# Lightweight security mechanism for PSTN-VoIP cooperation

Wojciech Mazurczyk[1], Zbigniew Kotulski[1,2]
[1] Warsaw University of Technology, Faculty of Electronics and Information
Technology, Institute of Telecommunications
{W.Mazurczyk Z.Kotulski}@tele.pw.edu.pl
[2] Polish Academy of Sciences, Institute of Fundamental Technological Research
zkotulsk@ippt.gov.pl

**Abstract**
In this paper we describe a new, lightweight security mechanism for PSTN-VoIP cooperation that is based on two information hiding techniques: digital watermarking and steganography. Proposed scheme is especially suitable for PSTN-IP-PSTN (toll-by-passing) scenario which nowadays is very popular application of IP Telephony systems. With the use of this mechanism we authenticate end-to-end transmitted voice between PSTN users. Additionally we improve IP part traffic security (both media stream and VoIP signalling messages). Exemplary scenario is presented for SIP signalling protocol along with SIP-T extension and H.248/Megaco protocol.

## 1. INTRODUCTION

Currently, PSTN (Public Switched Telephone Network) voice services are characterized by three important parameters: good voice quality, high reliability and security. We are able to provide critical services such as emergency numbers or federal agencies with ability for lawful intercept. All of these issues should be addressed before VoIP (Voice over Internet Protocol) systems are deployed on mass scale and the security become the most critical area [5]. With the deployment of IP Telephony accelerating, comes the increasing need to find ways to effectively secure it and to make it as secure as PSTN. That is why in IP network environment, especially for real-time services like VoIP, it has become desirable to develop security mechanisms which efficiently utilize available resources.

Current IP network's security solutions are usually based on deployment of a number of security devices and applications to protect and monitor networks such as: firewalls, Intrusion Detection (Prevention) Systems (IDS/IPS), Virtual Private Networks (VPN), authentication services and anti-virus software etc. Paradoxically, VoIP is highly sensitive to delay, packet loss and jitter, making these security mechanisms inadequate. Another challenge is that there must be a cooperation between security measures for IP Telephony. The lack of it is making protecting IP Telephony service ineffective from more sophisticated attacks or internal threats. So if we consider sending the calls in PSTN-IP-PSTN scenario, which is nowadays one of the most popular application for IP Telephony systems, we must emphasize that the IP part there is the weakest one from the security point of view.

That is why to adjust those two different environments we propose a new, lightweight security mechanism for PSTN-VoIP cooperation based on digital watermarking and steganography, so we can establish a reasonable tradeoff between security and performance even for low-bandwidth environment.

## 2. MOTIVATION

First we would like to emphasize that it is not enough to protect the media stream exchanged between the calling parties. This leads to a situation, in which the two entities are able to talk each other in a secure manner, but they are unable to initiate a call because of an attacker's actions on the signalling protocol. Usually, the users do not realize how much the weak security of signalling protocol, or its absence, can impact the security of the whole system. They often only demand their conversations be confidential and not revealed to the third-party.

As we mentioned in Section 1 the real problem for cooperation between legacy and new voice systems is securing IP Telephony part of PSTN-IP-PSTN connections. We should "seal" VoIP security gaps, so it will not affect the PSTN part. That is why, in this paper, we are proposing a different approach to the PSTN-VoIP security, based on digital watermarking. Our motivation to increase security for PSTN-VoIP interconnections is based on the following facts:
- There is a need for security mechanism for PSTN-IP-PSTN scenario to adjust security of the IP part to PSTN part,
- There is a need for low-power computing and low-bandwidth consuming mechanism for real-time services like VoIP to ensure certain level of security without affecting performance,
– There are no standardized security solutions for interworking between different signalling protocols, e.g. ISUP and SIP or even SIP and H.323. Every signalling protocol use disjoint set of security mechanisms. In PSTN-IP-PSTN scenario we must ensure that PSTN signaling messages as well as conversation will be transmitted in a secure manner during traveling through the IP environment,

– It is hard to secure VoIP system based on one VoIP signaling protocol [5] and providing secure interworking in PSTN-VoIP scenario is much more complex,
– VoIP security is still evolving. Lately, a large number of new security mechanisms was standardized for IP Telephony especially those for securing signalling messages but they often have certain disadvantages (like the big overhead in S/MIME for SIP or the use of TLS only for TCP, whereas traffic is carried mostly by UDP). So the process is not finished and still it is time for new solutions and ideas especially for securing PSTN-VoIP interworking,

The rest of the paper is organized as follows. In Section 3 general idea of proposed solution is presented. Then in Section 4 the mechanism's operations in IP network and PSTN part of the PSTN-IP-PSTN scenario are presented. Next in Section 5 security services that we gain with the use of this solution are outlined. Finally, we end with conclusions in Section 6.

## 3. GENERAL IDEA OF PROPOSED SOLUTION

We proposed using digital watermarking and network steganography techniques to improve VoIP security in [1], [2] and [3]. In [1] we used digital watermarking to authenticate VoIP conversation as well as messages of the signalling protocol that IP Telephony is based on. In [2] we used covert channel created in voice to alternate RTCP protocol functionality and additionally to provide security of VoIP conversation. In this solution we combined both information hiding techniques mentioned earlier. In [3] general description of protocol that uses multipurpose covert channel in VoIP transmission was defined. Network steganography was used to transmit the header and digital watermarking was used to transmit payload of protocol's PDU (Protocol Data Unit).

Here we propose to combine proposed solutions to improve security in PSTN-IP-PSTN scenario. The main idea is to authenticate conversation end-to-end (between PSTN users) with the watermark embedded into the voice and additionally use network steganography and digital watermarking in the in IP environment to ensure VoIP signaling protocol and ISUP (PSTN signalling protocol) security. So in this solution we will embed two watermarks into the voice stream: one at the PSTN endpoint, second at the entrance to the IP part of the network. The watermarking technique that is used must allow to embed/extract two different watermarks without distortion or destroying the existing ones. That means that we will be watermarking already watermarked voice stream.

The solution general idea is also presented in Fig. 1.

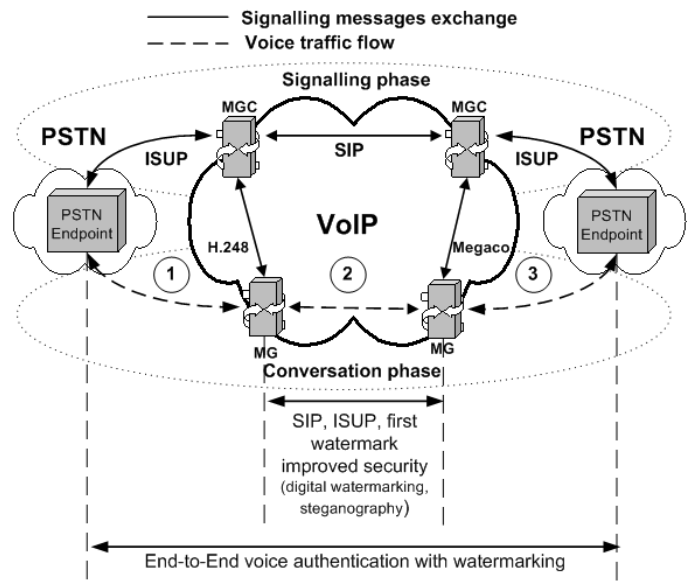

**Fig. 1.** Proposed solution general operation idea

We will reference and describe the above figure in the next sections. In next two sections we will describe in details proposed solution for IP then PSTN part of the PSTN-IP-PSTN scenario.

## 4. THE IP PART OF PSTN-IP-PSTN SCENARIO

In this section we will focus only on the part of Fig.1 that is marked with ②. We will demonstrate the IP part of the solution based on SIP protocol [6] and its extension SIP-T [7], because it is nowadays one of the most popular signalling protocol for IP Telephony. SIP-T (SIP for Telephones) is an extension to SIP protocol that allows it to be used for ISUP call setup between SS7-based public switched telephone and SIP-based IP telephony networks. SIP-T carries an ISUP message payload in the body of a SIP message. Transporting ISUP in SIP bodies may provide opportunities for abuse, fraud, and privacy concerns, especially when SIP-T requests can be generated, inspected or modified during the travel in IP environment. The standard proposes that ISUP MIME bodies should be secured (preferably with S/MIME) to alleviate this concern. But MIME (as well as S/MIME) has certain disadvantages for real-time services as stated in [5].

In this paper MGC (Media Gateway Controller) and MG (Media Gateway) concept is also used for VoIP-PSTN interworking. MG converts media provided in one type of network to the format required in another type of network. MGC controls the parts of the call state that pertain to connection control for media channels in a MG. Interactions between MGC and MG are described in IETF and ITU joint standard H.248/Megaco [8].

Our solution will be showed on the PSTN-IP-PSTN scenario characteristic for SIP that is defined in [7]: SIP bridging. It means that the IP network (with SIP as a signalling protocol for VoIP) will be used only as a transport network – both caller and callee are in PSTN. This is a popular scenario for today's VoIP providers, which is also called toll-by-passing.

**4.1. DIGITAL WATERMARKING SCHEME MODIFICATIONS AND TOKEN CREATION**

Let us present the watermarking scheme (originally proposed in [1]) that will be also used here. Every Media Gateway (MG) is equipped with the functional block called Pre-processing Stage (PPS), see Fig. 2, which is responsible for preparing data before the watermark embedding stage. Its elements do the following.

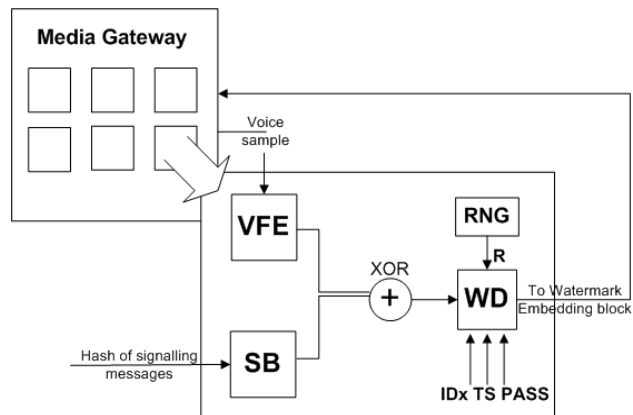

**Fig. 2**. Pre-processing stage block (PPS) in Media Gateway (MG)

**SB** (Signalling Message Hash Buffer): stores hashes of signalling messages (or those fields that are not changed during transmission) from the first phase of the call provided from MGC (for ISUP and SIP protocols).

**VFE** (Voice Feature Extractor): provides characteristic features (VF) of the original voice that is sent from PSTN network. Afterwards a hash can be also performed on this value.

**RNG** (Randomizer): generates a random number R. We use it to provide a unique set of data for every embedded watermark.

**WD** (Watermarking Data): in this block the input data is concatenated with R, IDX (unique, global identifier of one party of the connection), PASS (the password known to the both parties) and TS (time stamp).

The embedded watermark formed as described above we will call a **token**. At the entrance and at the end of the IP network part of PSTN-IP-PSTN scenario the received token with a locally calculated, appropriate one. Fig. 3 shows, how the algorithm works for SIP protocol with SIP-T extension (to protect ISUP messages that are contained within SIP messages). It is SIP bridging scenario: the caller and callee are in PSTN and their communication is sent through IP network (MG sends RTP packets to the other Media Gateway).

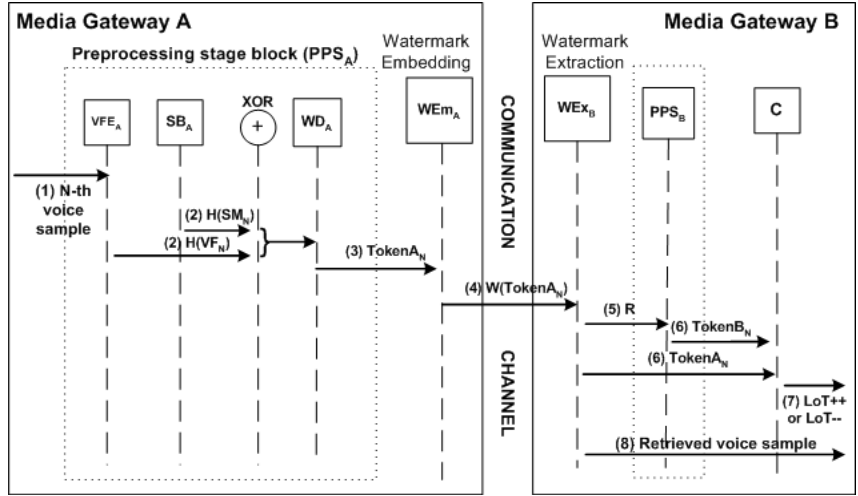
**Fig. 3.** Proposed authentication solution operation for MG-MG connection

In this situation, the values of the tokens $A_N$ and $B_N$ are:

$$TokenA_N = TokenB_N = H\left( (H(SM_N) \oplus H(VF_N)) \| \begin{pmatrix} TS \\ PASS \\ ID_A \end{pmatrix} \| R \right) \|$$

$PPS_B$ block is analogous to $PPS_A$ in functioning. Additionally, we assume that in the initial signalling phase some of the signalling messages, for both signaling protocols: ISUP and SIP ($SM_N$ means the N-th Signalling Message) were exchanged (and their hashes are stored in SB). In the second phase they are verified. **H** stands for the hash function and **W** for embedding of the digital watermark into audio. The algorithm works as follows:

  (1) When the conversation begins, the first voice sample enters $VFE_A$ block the feature of the voice sample ($VF_N$) is extracted (for the data integrity) and then the hash function is performed on the result.
  (2) The values from $SB_A$ and $VFE_A$ are then XORed (they have the same length). Afterwards, the result is sent to $WD_A$ block, in which **TokenA** is created together with the other parameters like: the randomizer value (R), shared password (PASS), global identifier of A ($ID_A$) and, optionally, the time stamp (TS).
  (3) TokenA is sent to the watermark embedding function and the information, that it contains, is saved there in the caller's voice. Then, the data stream formed in this way is sent through the communication channel.
  (4) Before the voice from MG A can be processed by MG B, the watermark is extracted and sent to the comparator (C) on the receiver's side.
  (5) From the extracted token, the randomizer value (R) is sent to the analogous $PPS_B$ block. In this block some pre-processing had taken place. If we have the R value, then we can compute **TokenB**. It should be equal to TokenA, if the transmission had not been tampered. The result is sent to the comparator (C).
  (6) In C both token values are compared.
  (7) If TokenA=TokenB, the special parameter LoT (Level of Trust) increases (it reflects a number of correct tokens received). Otherwise it decreases. Then, depending on LoT value the decision is made (by MGC), whether the call should be continued or broken down. Computation the LoT parameter is described in details in [1, 2].
  (8) If MGC decides that the call can be continued, the voice sample finally can be converted to PSTN format and directed to callee.

Thus, the authentication and integrity process depends on exchanging the security tokens and their comparison with the ones locally calculated at the receiver. It is essential to deal properly with every signalling messages exchanged during the connection. Those messages should not influence the call until they are authenticated and their integrity is verified. Authentication of messages, which are used to terminate VoIP conversations have to be treated the same as a normal message that comes during the call. Normally, the media channels are terminated, upon receiving this message. In our scheme it is vital to retain RTP flow until those messages are authenticated.

As we mentioned earlier MGC and MG are vital for PSTN-VoIP interconnections. H.248/Megaco protocol was created for this purpose. Based on [8] we propose that MG uses *Notify* message to inform MGC that calculated and received tokens are the same or not. MGC should make a decision to continue or break a connection. Almost all the calculations are passed to MG, but MGC must perform a hash function on signalling messages (or on parts of them) that were exchanged during the first phase of the call (and later). Then MGC sends those hash values to be stored in MG.

**4.2. NETWORK STEGANOGRAPHY TO DISTINGUISH PARAMETERS**
Besides using digital watermarking in the IP part we will use network steganography to be able to send additional data as described in details in [2, 3]. In this way we will create a multipurpose covert channel. For IP network part we will use it generally for post authentication (as described in [3]). The general protocol overview is presented in the Fig. 4:

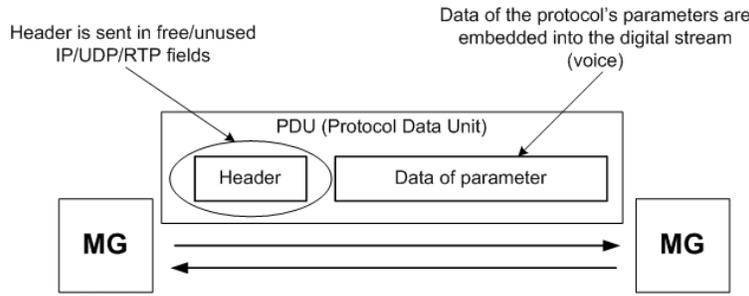

**Fig. 4.** Using network steganography to distinguish parameters to send

The size of the header of the PDU (Protocol Data Unit) shown in the Fig. 4 is small. In [3] it contained 6 bits that were entered in the unused IP/UDP/RTP fields in the certain way known to both sides of the communication. Moreover PDU can have one of two payload types: security or informational. How the informational parameters can be used freely to carry any data that is needed in the communication process e.g. they can be used to alternate RTCP protocol functionality as we showed in [2]. But more important here are the security payloads. That means that PDU contains certain authentication and/or integrity information that should be verified after its extraction form voice. Two kinds of security payloads are available, first is used to provide authentication and integrity of the voice, its source and signalling protocol messages (e.g. a token as described in 4.1). Second's role is to provide post authentication for protocol parameters that were send earlier (both security and informational). That gives greater security of the whole digital stream and transmission. The general idea of its calculation is presented in Figure 5.

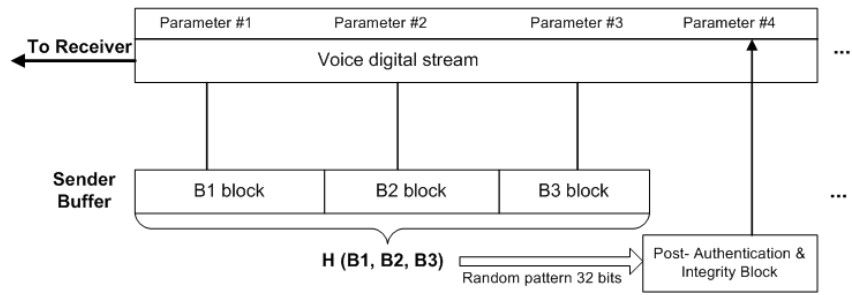

**Fig. 5.** Example of the post authentication of transmitted parameters

As we see in the Fig. 5 there is a chaining between the transmitted parameters. Every n-th parameter is used for post authentication of n-1 parameters that were send earlier (For Fig. 4. n=4).

**5. THE PSTN PART OF PSTN-IP-PSTN SCENARIO**
To fulfill the requirements for proposed solution, special enhanced PSTN endpoint must be used. It must be capable of embedding/extraction watermark. We do not dictate watermarking technique that will be used in PSTN endpoint to embed watermark into voice – although it can be digital or analog watermarking scheme. The important thing is that the watermark that is created in the voice must survive AD/DA operation and embedding of different watermarks. For PSTN endpoint two modes are available:
- It can embed watermark using analog watermarking technique. In this case the watermark must be characteristic for particular user. The watermark created must also reflect special features of conversation send as it was shown in case of IP part of PSTN-IP-PSTN connection.

- It can convert analog voice into digital domain, embed analogous watermark as presented in Section III and then convert it again to analog signal and transmit through PSTN network.

So at the endpoint the watermark is embedded and it is verified at the other end of PSTN-IP-PSTN connection (①and ③ in Fig.1). During the retrieving process the other PSTN user verifies only the watermark embedded by the caller. The other data (added during the traveling through the IP network part) is ignored.

## 5. SECURITY SERVICES FOR PSTN-IP-PSTN SCENARIO

Proposed solution as shown in Fig. 1, that uses watermarking technique and network steganography greatly improves security for PSTN-IP-PSTN scenario. In PSTN part it provides (① and ③ in Fig.1):
- **Authentication of the data source** (one can be sure of the identity of the caller),
- **Data authentication and integrity** (one can be sure that the audio comes from the caller and it has not been tampered).

Using our scheme in the IP network part (In Fig. 1 marked with ②) provides, besides mentioned above, following security services:
- **Authentication of the signalling messages for SIP and ISUP** (one can prove that the caller is the source of the signalling messages that were exchanged during the signalling phase of the call),
- **Signalling messages integrity** (one knows that the signalling messages were not modified during the transmission through the communication channel),

Additionally security provided in the IP part is enhanced with the use of post authentication parameter as described in 4.2.

As we see with the use of proposed mechanism we can provide certain security services for voice communication between VoIP system and PSTN. Moreover security services provided in PSTN part are guaranteed in the end-to-end manner.

## 6. CONCLUSIONS

The proposed lightweight PSTN-VoIP secure cooperation solution defines a useful mechanism that provides authentication and integrity of the voice traffic (both conversation and signaling messages) along the communication path. It takes advantage of two information hiding techniques: watermarking and network steganography, so it does not consume any user bandwidth.

As showed in this paper it is suitable especially for PSTN-IP-PSTN scenario and it adjusts IP network part security to the PSTN level. We are able to verify authentication of both media streams as well as the signalling protocols used (ISUP, SIP). In this way we gain one of the first security mechanisms that works in the heterogeneous environment for voice communication